\documentclass[bibnotes,amsmath,amssymb,showpacs,floatfix,superscriptaddress,reprint]{revtex4-1}
\usepackage{amsmath,empheq}
\usepackage{amsfonts}
\usepackage{amssymb}
\usepackage{amsxtra}
\usepackage{mathtools}
\usepackage{xcolor}
\usepackage{graphicx}
\usepackage{subfigure}
\usepackage{dcolumn}
\usepackage{mathrsfs}
\usepackage{float}
\usepackage{bm}
\usepackage[breaklinks=true,colorlinks,citecolor=blue,linkcolor=blue,urlcolor=blue]{hyperref}

\DeclareMathAlphabet{\bi}{OML}{cmm}{b}{it}
\def\be{\begin{equation}}
\def\ee{\end{equation}}
\def\bearr{\begin{eqnarray}}
\def\eearr{\end{eqnarray}}

\begin{document}
\title{Topological nonlocal operations on toroidal flux qubits}
\author{Adel Ali}
\email{adel_ali@tamu.edu}
\affiliation{Department of Physics and Astronomy, Texas A\&M University, College Station, TX, 77843 USA}
\author{Alexey Belyanin}
\email{belyanin@tamu.edu}
\affiliation{Department of Physics and Astronomy, Texas A\&M University, College Station, TX, 77843 USA}

\date{\today}

\begin{abstract}
We propose a conceptual model of a toroidal flux qubit, which consists of a quantized toroidal magnetic flux coupled to a charged particle on a quantum ring through field-free interaction. Scaling the system to two or more flux qubits results in emergent field-free coupling between them. We show that the topological and nonlocal aspects of this system can have profound applications in quantum information. We illustrate it with examples of nonlocal operations on these flux qubits which are protected from environmental noise, including creating entanglement and ``teleporting'' excitation energy between the flux qubits. 
\end{abstract}

%
\maketitle
{\it\underline{Introduction}:}
The quest for scalable quantum computing is believed to be contingent on reducing decoherence and errors \cite{Siddiqi_2021}. Topological quantum computing is an attempt to do so by encoding the quantum information in a many-body system where the local errors cannot destroy the encoded information. Despite the mathematical beauty of this idea, finding materials that realize non-abelian anyons has not been unambiguously confirmed \cite{stern2013topological}.
Furthermore, storing and processing quantum information requires two seemingly contradictory processes: complete isolation of the system from the environment, and precise control of the system's degrees of freedom. One example of a well isolated qubit candidate is a toroidal flux qubit or fluxon, which is defined as a volume of enclosed magnetic flux lines whose outer surface is topologically equivalent to a torus. 
As in the case of normal flux qubits, the persistent toroidal current generating this magnetic flux can be in a superposition \cite{Orlando_1999}.
Fluxons can be isolated from the environmental noise such as flux noise, charge noise, 1/f noise, external electromagnetic fields, etc. 
\cite{nori, noise}. 
This raises the question: how to do operations on its quantum state if it is isolated from the environment? 
We propose to solve this problem by coupling the fluxon with a charged particle living in a ring that encloses the flux lines of the fluxon, or the quantum ring (QR) \cite{korea,Vinasco,PhysRevB.64.045327,Dauber_2017,Fuhrer_2001}. The fluxon-electron interaction is mediated by the coupling of the electron with the gauge field of the fluxon, i.e., essentially the field-quantized version of the Aharonov-Bohm effect. Any noise charge must close a loop around the fluxon to get coupled to it. The environmental noise charge fluctuations do not affect the system unless they make coherently a trajectory that encloses the flux with a non-zero winding number. 

The Aharonov-Bohm (AB) phase has been studied in a context where the gauge vector potential of a classical magnetic field affects nonlocally a charged particle's phase for scattered states and the energies of the bound states \cite{review}. However, this field has not been quantized. We propose and study the {\it quantized flux} scenario. This means that the magnetic flux itself can be in a superposition of different quantum  states.  

The AB effect has two characteristic features: its topological nature which means that it is protected from deformations of the flux lines and the charged particle path, and its nonlocal nature which means that the flux can affect the phase of a distant particle. These two features are of great significance for fundamental physics and, as we show below, may lead to unique quantum information applications. In this Letter, we analyze the quantized flux version of the AB system and show that the joint quantum state of the particle and fluxon can be manipulated by the external electromagnetic (EM) fields acting only on the electron degrees of freedom; hence we can deterministically control the fluxon state. Moreover, 
we show that for an electron coupled to two fluxons that are separated in space and isolated from each other and the environment, entanglement can be created between them, and excitation energy can "teleport" from one to the other. This reasoning is scalable to a chain of coupled fluxons where it simulates interesting spin chain systems with nonlocal coupling. 

The proposed system is conceptually simple enough for a semi-analytic treatment and at the same time, it presents an example of synergy of several fundamental physics phenomena and models, such as the AB effect, flux qubits, spin chains, and tunable Ising models, as well as their application to quantum information.

%
{\it\underline{AB phase topological protection}}:
Consider a QR that surrounds a volume with a magnetic flux trapped inside, as sketched in Fig. 1a. The eigenvalues and eigenvectors of the system are invariant under deformations of the charge path (given that this deformation is approximately length preserving), deformations of the flux lines, or simultaneous deformations of the charge path and the closed flux lines \cite{deformation}. Indeed, in the semiclassical Bohr–Sommerfeld quantization condition, 
$$
\hbar \oint d \mathbf{r} \cdot \mathbf{k}-e \oint d \mathbf{r} \cdot \mathbf{A}=2 \pi \hbar n
$$
the second term is clearly invariant under the deformation of the closed flux lines. For length preserving deformations the first term is also invariant. Hence, the allowed momenta and energies 
are invariant and these systems have topological protection. That may address the design and fabrication challenges such as uniformity and reproducibility across qubits, disorder, and fabrication defects.

\begin{figure}[t]
\includegraphics[width=7cm]{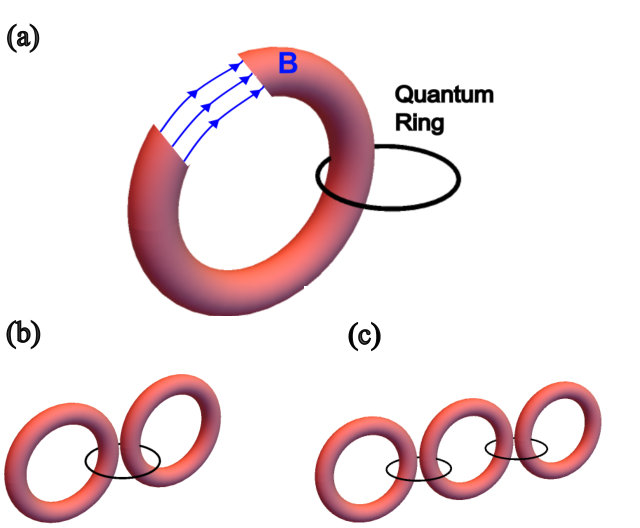}
\centering
\label{one}
\caption{(a) Schematic of a fluxon coupled to an electron on a quantum ring. The magnetic field is confined inside the toroid and the electron on the ring is coupled to the vector potential only.
(b) two fluxons coupled nonlocally through the electron on a quantum ring. 
(c) A chain of coupled fluxons. }
\end{figure}

{\it\underline{Fluxon-electron coupling}}:
The Hamiltonian for a massive charged particle in a field-free region outside of an infinite solenoid is given by 
$$
\hat{H} = \frac{1}{2m_e} (\hat{p}_{\varphi}-e \hat{A}_{\varphi})^{2}; \, A_{\varphi}=\frac{\Phi}{2 \pi r} 
$$
and $\varphi$ is an angular coordinate along the QR which changes from 0 to $2\pi$. We quantize the enclosed flux $\Phi$ by demanding that $\Phi=\hat{n} \varphi_{0}$ where $\hat{n}$ is a number operator, and $\varphi_{0} = h/2e$ is the flux quantum.

For all examples below we will limit the fluxon excitations to the two lowest states with $n$ equal to 0 or 1 in order to have a two-state flux qubit, which is the simplest and most studied case for applications. In the ``physical'' basis $|n=0\rangle$, $|n=1\rangle$ the fluxon Hamiltonian is $\Delta \sigma_{x}$ where $\Delta$ is the transition energy and $\sigma_x$ is a Pauli matrix. Note that it is often written as $\Delta \sigma_{z}$, after performing the Hadamard transformation \cite{nori}.  Furthermore, we will start with the $1D$ motion of a particle on a ring. The case of a finite-width 2D annulus geometry is qualitatively the same, unless we want to utilize the energy levels related to transverse quantization which happens at higher energy scale. The resulting fluxon-electron Hamiltonian is  
\begin{equation}
\hat{H} = \frac{1}{2m_e} \left(\hat{p}_{\varphi}-\frac{e \varphi_{0}  \hat{n}}{2 \pi r}\right)^{2}+ \Delta \sigma_{x}
\label{ham1}
\end{equation}
We take the basis of the Hilbert space to be 
$  \left|\psi_{m n}\rangle =\right| m\rangle \otimes|n\rangle $ where 
$$ \left\langle\varphi \mid m\right\rangle=\frac{e^{-i m \varphi}}{\sqrt{2 \pi r}} \text {, and }    \hat{n}|n\rangle=n|n\rangle. $$
Using the ansatz $\left|\psi_{m}\right\rangle=|m\rangle \otimes(\alpha|0\rangle+\beta|1\rangle)$
to solve the time-independent Schroedinger equation $\hat{H}|\psi_m\rangle=E_m|\psi_m\rangle$, we obtain 
\begin{equation}
\begin{aligned}
\hat{H}\left|\psi_{m}\right\rangle  =\frac{\hbar^{2}}{2 m_e r^{2}}\left[|m\rangle \otimes(\alpha m^{2} |0\rangle)+((m-1)^{2}\beta  |1\rangle)\right]\\ 
+\Delta[|m\rangle \otimes \alpha|1\rangle+|m\rangle \otimes \beta|0\rangle] |\psi_{m}\rangle.
\end{aligned}
\nonumber \end{equation}

Setting the energy scale $a = \frac{\hbar^{2}}{2 m_e r^{2}} \equiv 1$, so that $\Delta$ is now in units of $a$ and simplifying, we arrive at 
$$
\hat{H}\left|\psi_{m}\right\rangle = |m\rangle \otimes\left[\left(\alpha m^{2}+\Delta\beta\right)|0\rangle+\left(\beta(m-1)^{2}+\Delta \alpha\right)|1\rangle \right] $$
$$
=  E_{m}[|m\rangle \otimes(\alpha|0\rangle+\beta|1\rangle)]. $$
This can be written as 
\begin{equation} 
{\left[\begin{array}{cc}
m^{2} & \Delta \\
\Delta & (m-1)^{2}
\end{array}\right]\left[\begin{array}{l}
\alpha \\
\beta
\end{array}\right]=E_{m}\left[\begin{array}{l}
\alpha \\
\beta
\end{array}\right] }
\label{eigen}
\end{equation}
Solving this eigenvalue problem gives
\begin{equation}
E_{m}^{ \pm}=\frac{1}{2}\left(1-2 m+2 m^{2} \pm \sqrt{(1-2 m)^{2}+4 \Delta^{2}}\right). 
\label{energies}
\end{equation} 
The ansatz is a product state of $m$ and $n$ states, so the stationary states of the joint system have the fluxon part given by the eigenstates of the effective Hamiltonian matrix on the left-hand side of Eq.~(\ref{eigen}). This matrix can be written as  $h_{0} I + \Vec{h}.\vec{\sigma}$ where $\vec{\sigma}=(\sigma_{x},\sigma_{y},\sigma_{z})$,  
$\vec{h}=(\Delta,0,m-\frac{1}{2} )$, and $h_{0}=m^{2}-m+1$.   
In spin-1/2 terminology, $\Delta$  is the $x$-component of the effective magnetic field, $h_{3}$ is its $z$-component, and $h_{0}$ is a total energy shift. So changing the quantum number $m$ (for example by the EM field) will result in changing the effective Hamiltonian of the fluxon. 

The eigenvectors describing the fluxon part of the eigenstates, namely $\alpha|0\rangle+\beta|1\rangle$, are shown as arrows on the Bloch sphere in Fig. 2(a), (c), and (d) for different values of $\Delta$, whereas the eigenenergies from Eq.~(\ref{energies}) are shown in Fig.~2(b). If $\Delta \ll 1$, the eigenvectors are concentrated around the poles of the Bloch sphere as in Fig.~2(d) because the  z-component will dominate. As $\Delta$ increases, the states initially condensed on the poles will start to spread out as shown in Fig.~2(a) for $\Delta = 5$ and (c) for $\Delta \gg 1$. In particular, the states with $m$ of the same order as $\Delta$ will move towards the x-axis of the Bloch sphere. One notices from Fig.~2(a) and (b) that even when $m=0$, i.e., the electron has zero angular momentum as is the case for state $|g_1\rangle$, the value of $h_{3}=-1/2$ is still non-zero and the electron still affects the state of the fluxon. This is analogous to the vacuum effects in the Jaynes–Cummings model.
 


\begin{figure*}[t]
\includegraphics[width=\linewidth]{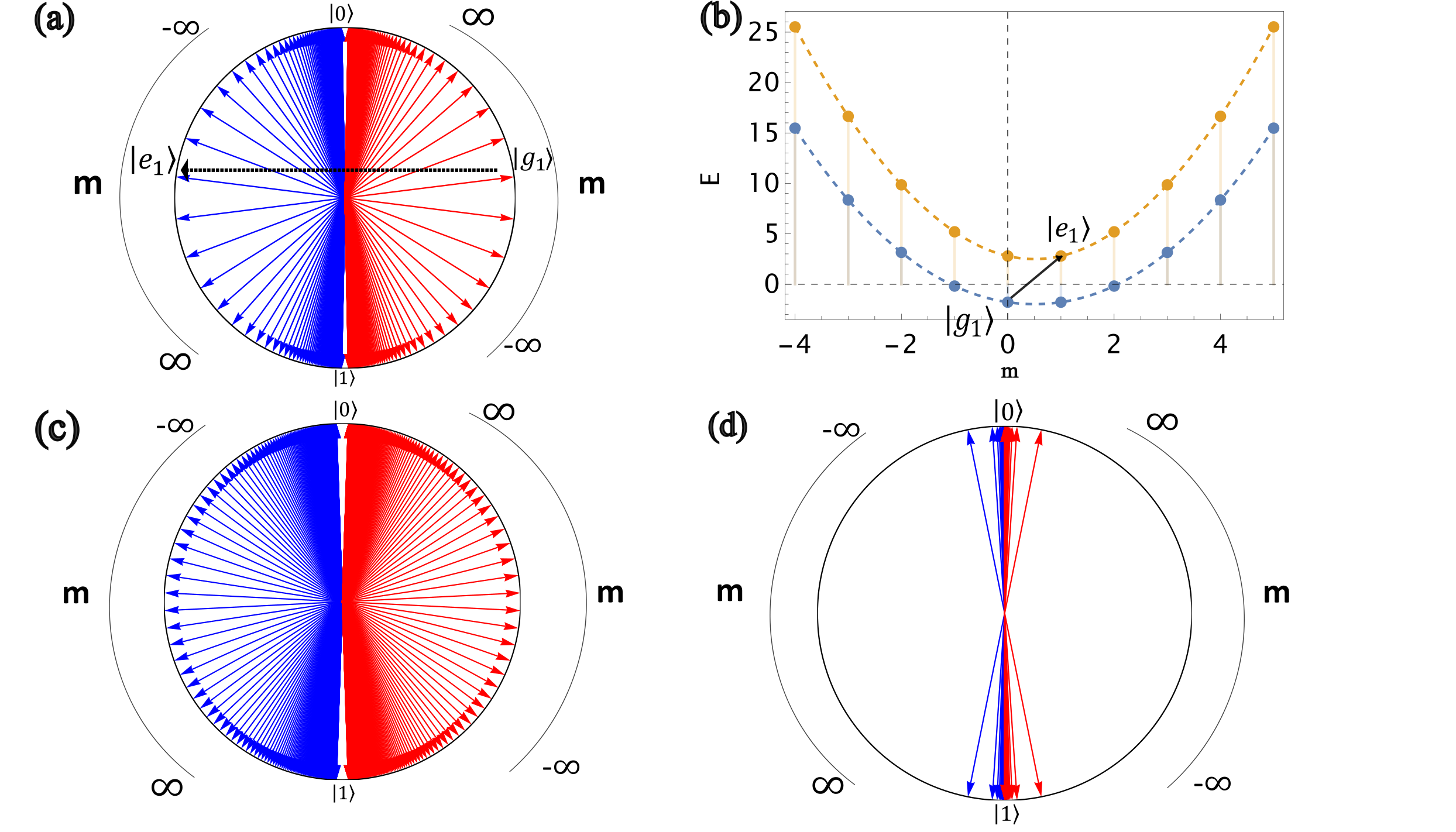}
\centering
\caption{(a) Plot of the eigenstates of the fluxon for $\Delta=5$ on a section of a Bloch sphere, where each arrow represents a state that has a different $m$ number. The red arrows on the right and blue arrows on the left correspond, respectively, to the lower and  the upper energy bands shown in part (b). The horizontal arrow shows an example of an electric-dipole allowed transition.  (b) The energy spectrum of the states in (a) with the transition from part (a) shown with an arrow. (c) Same as part (a) but for $\Delta \gg 1$. (d) Same as part (a) but for $\Delta \ll 1$. }
\label{fg2}
\end{figure*}

{\it\underline{Operations}}~:
Can we use the gauge-field nonlocal coupling between the electron and the fluxon to do quantum operations on the fluxon(s)? 
The fluxons are completely shielded from any EM field and we have access only to the electron on the QR degrees of freedom. The question is significant for fundamental physics, e.g., within the debate on the observable effects of the gauge field potential. This nonlocal coupling can be used to transfer energy and information between the electron and the fluxon, and mediate interaction between fluxons. It also has important implications for quantum information technologies which rely on the fact that a charged particle that encloses the fluxon can interact nonlocally with the fluxon while the fluxon is isolated from virtually everything else in the environment.

Consider as an example the transitions between the electron $m$-states caused by an EM field and their effect on the fluxon state. The external EM field will couple only to the electron moments. In the electric dipole approximation, the selection rules are $\Delta m=\pm 1$.  A simple operation on the fluxon would be to excite the joint system to a higher state, as shown in Fig.~\ref{fg2}(b). This will result in rotating the fluxon state on the Bloch sphere as indicated in Fig.~\ref{fg2}(a). Note that each eigenstate of the system is a product state,  so after the excitation, even if the ring were somehow ``cut'', the x-basis measurement of the fluxon would not change, as the fluxon state would rotate around the x-axis. Since the fluxon is well isolated from the environment, it can be used to store quantum states.  

Another set of operations to manipulate the state of the fluxon is to add a potential to the electron on the ring, for example, by applying lateral electric field(s). Consider the lower two $m$ states only, so that the Hamiltonian can be mapped to a two-qubit Ising model: $ \hat{H}_I = (- \sigma^{(e)}_{z} \sigma^{(f)}_{z}+ I)  +\Delta \sigma_{x}^{(f)} + g_1(t)  \sigma_{x}^{(e)}+g_2(t)  \sigma_{y}^{(e)}$. Here superscripts $(e)$ and $(f)$ denote the operators acting only on electron and fluxon degrees of freedom, respectively. Then geometric phase gates can be performed on the fluxon \cite{ising} by adiabatically changing $g_{1,2}(t)$. 

Creating entanglement between the electron state and the fluxon has interesting implications in the context of nonlocal gauge effects. Indeed, consider an initial state of the system to be a product state of the electron and the fluxon, so that there is no correlation between the measurements done on both. 
To illustrate the idea of nonlocal effects with the same simple model, 
we take the same Hamiltonian as in Eq.~(\ref{ham1}) but with an arbitrary coupling of the electron to an external field: 
\begin{equation}
\hat{H}=\frac{1}{2m_e} \left(\hat{p}_{\varphi}^{(e)}-\frac{e \varphi_{0}  \hat{n}^{(f)}}{2 \pi r}\right)^{2}+\Delta \sigma_{x}^{(f)}+\hat{H_{e}},
\label{ham3}
\end{equation}
where $H_{e}=g(t)  \sigma_{x}^{(e)}$.  The operator $\hat{H_{e}}$ is acting only on the electron subspace and can describe, e.g., the interaction of the electron dipole moment with an EM field. 
In the basis of $|m,n\rangle$ for the lowest energy states of the uncoupled system, $\{|00\rangle,|01\rangle,|10\rangle,|11\rangle\}$, the matrix elements of the Hamiltonian (\ref{ham3}) are 
\begin{equation}
H=\left[\begin{array}{cccc}
0 & \Delta & g & 0 \\
\Delta & 1 & 0 & g \\
g & 0 & 1& \Delta \\
0 & g & \Delta & 0
\end{array}\right]
\label{ham4}
\end{equation}

According to our scenario, at $t \leq 0$ we have $g = 0$ and the eigenstates of the Hamiltonian (\ref{ham4}) are product states of the $|m\rangle$ and $|n\rangle$ basis states. At $t = 0$ the interaction term $H_{e}$, which models the dipole coupling of the electron with an external EM field, is turned on. Hence, driving the electron will  result in inducing the coupling coefficient $g$ in (\ref{ham4}). Accordingly, the time evolution will result in changing the reduced density matrix of the fluxon. One can easily check numerically that the expectation value of the energy and von Neumann entanglement entropy of the fluxon will increase due to the coupling. This implies that the exchange of information and energy between the two subsystems can happen nonlocally mediated by the gauge field potential only. 


{\it\underline{Nonlocal coupling and operations on  fluxons}}:
The system can be scaled up to include more complicated operations.  Consider a setup that involves two of such fluxons, as in Fig.~1(b). 
The Hamiltonian is now
\begin{equation}
    \hat{H}=\left(\hat{p}_{\varphi}-\frac{e \varphi_{0} \hat{n}_{1}}{2 \pi}+\frac{e \varphi_{0}}{2 \pi} \hat{n}_{2}\right)^{2}+\Delta \sigma_{x}^{(f1)} +\Delta \sigma_{x}^{(f2)}.
    \label{ham5}
    \end{equation}
Here $(f1)$ and $(f2)$ denote operators acting on the degrees of freedom of fluxon 1 and 2. We assumed the configuration in which the two fluxons have opposite directions; hence the opposite signs in parentheses in Eq.~(\ref{ham5}).
For the eigenstates of the system, we use the ansatz 
$$\left|\psi_{m}\right\rangle=|m\rangle \otimes\left( a_{1}|00\rangle+a_{2}|01\rangle+a_{3}|10\rangle+a_{4}|1 1\rangle \right), $$
and the effective Hamiltonian becomes
$$\left[\begin{array}{cccc}
m^{2} & \Delta & \Delta & 0 \\
\Delta & (m-1)^{2} & 0 & \Delta \\
\Delta & 0 & (m-1)^{2} & \Delta \\
0 & \Delta & \Delta & m^{2}
\end{array}\right]$$

Its  eigenenergies are  
$E_{m}^{(1)}=(m-1)^{2}, E_{m}^{(2)}=m^{2}$\\ $E_{m}^{(3,4)}=\frac{1}{2}\left(1-2 m+2 m^{2} \pm \sqrt{(1-2m)^2+16 \Delta ^2}\right).$

\begin{figure}[t]
\includegraphics[width=7cm]{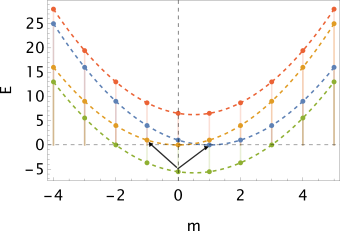}
\centering
\caption{The eigenenergies of the two-fluxon system plotted as a function of $m$ for $\Delta=3$. The black arrows represent electric dipole-allowed transitions from the ground state which is a product state for the fluxons $|n_1\rangle \otimes |n_2\rangle$,  to excited states in the second and third bands which are entangled Bell states.  }
\end{figure}

They are plotted in Fig.~3. The salient feature of the system is that the fluxon parts of the corresponding eigenstates are entangled states of the fluxons. For example, $|\Phi^{-}\rangle=\frac{|00\rangle - |11\rangle}{\sqrt{2}}$ is the fluxon part of the eigenstate of the entire $E_{m}^{(1)}$ band, and $|\Psi^{-}\rangle=\frac{|01\rangle-|10\rangle}{\sqrt{2}}$ is the same for the $E_m^{(2)}$ band. All the states of these bands are still orthogonal because they have different quantum numbers $m$. These are maximally entangled Bell states of the fluxons. 
Therefore, a simple operation of exciting the system from the ground state $E^{(4)}_m$, which is a product state, to any state in the $E_m^{(1)}$ or $E_m^{(2)}$ bands as shown in Fig.~3  will nonlocally create a maximally entangled state between the two fluxons. This is a remarkable and potentially very useful feature of the system. 

The effective Hamiltonian given by the matrix above can be written as $$\hat{H}_{m}=h_{0} I + \Vec{h}.\vec{\sigma}^{(f1)}+\Vec{g}.\vec{\sigma}^{(f2)}+J_m  \sigma^{(f1)}_{z} \sigma^{(f2)}_{z}$$ where the nonzero coefficients are 
$$h_{0}=m^{2}, \; h_1=g_1=\Delta, \; h_{3}=g_{3}=m-1/2, \; J_m=2m-1. $$ 
The coupling coefficient $J_m$ between the two initially non-interacting fluxons emerged due to the coupling with the electron and describes electron-induced fluxon-fluxon interaction. Since $J_m$ is a function of $m$, it can be tuned and changed between antiferromagnetic for $m>0$ and ferromagnetic for $m<0$. Moreover, the coupling coefficient is topological, meaning that it does not change by deforming the QR or changing the distance between the two fluxons as long as the electron path encircles both fluxons. This is yet another unusual property of this system as opposed to the usual situation when the interaction strength falls off with the distance between the interacting parts of the system. 

\begin{figure}[t]
\includegraphics[width=8cm]{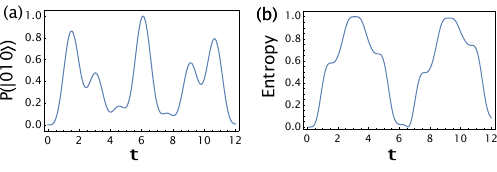}
\centering
\caption{Exchange of information and energy between the two fluxons. (a) for $\Delta=1$, the probability for measuring $|010\rangle$ is plotted as function of time normalized by $\hbar/a$. Around $t=6$ the excitation is completely transferred from fluxon 2 to fluxon 1.  (b) The von Neumann entanglement entropy of the fluxon as a function of time normalized by $\hbar/a$. }
\end{figure}

Another interesting feature of the coupled QR-fluxon models is that the energy can be transferred nonlocally between the two fluxons. To illustrate that, consider the initial state in the form $|\psi(0)\rangle=|m=0\rangle \otimes |n_{1}=0\rangle \otimes |n_{2}=1\rangle$. The time evolution of this state will keep $m$ fixed and results in rotation of the fluxons part of the wavefunction by the effective Hamiltonian above. Accordingly, after some time $\tau$, the state vector becomes $|\psi(\tau)\rangle=|0\rangle \otimes |1\rangle \otimes |0\rangle$. Fig.~4(a) shows the probability of finding the system with $n_1=1$. This means that the flux (energy) initially in fluxon 2 can ``teleport'' to  fluxon 1. This can be done with any value of $m$, in which case the coupling coefficient could be larger and $\tau$ would be smaller. However, the evolution starting from the state with $m=0$, i.e., the state with zero magnetic moment  is still most peculiar as it rules out any effect of the magnetic field produced by the electron.  

The models considered here can be straightforwardly scaled up or extended to include more operations. For example, more than one electron can be injected into a QR, which will lead to the occupation of several $m$-states and excited energy bands in Fig.~3. Furthermore, a chain of the fluxons can be created where each two consecutive fluxons are coupled to each other, as shown in Fig.~1(c). Following the same procedure as in the case of two fluxons, the effective Hamiltonian of the fluxons on the chain can be written as 
$$\hat{H}_{m}=\sum_{i} J_{m}\sigma^{(i)}_{z} \sigma^{(i+1)}_{z}+\Delta\sigma_{x}^{(i)}+h_{m}\sigma_{z}^{(i)}+h_{0} I. $$
This is a transverse field Ising Hamiltonian with tunable parameters. Hence, this system can be used as a quantum simulator \cite{nori2}. The two protocols discussed above can be applied for the chain setup. First, one can "teleport" energy by starting with only one fluxon excited, when the time evolution results in the excitation hopping between different sites of the chain. Second, entanglement of the fluxons on the chain can be created by entangling the fluxons on the chain pairwise consecutively by changing the $m$-state of an electron in the corresponding QR.  

Finally, all qualitative features of the models will be preserved if one replaces a nonrelativistic 1D  electron with any other type of the QR. For example, employing a carbon nanotube or a graphene ring will change the shape of the energy dispersion in Fig.~2(b) and 3 while all operations remain the same. The scale of the energy parameter $a$ defining the separation between the eigenstates will be of course affected. Depending on the radius and the type of the ring, one can expect the transition energies from the microwave to the mid-infrared range.





-------------------------

\bibliography{sample}
\end{document}